\documentclass[showpacs,aps,twocolumn]{revtex4}
\usepackage{epsfig}
\usepackage{graphicx}
\usepackage{amsmath,amssymb,amsfonts}
\usepackage{array}
\usepackage{url}
\usepackage{hyperref}
\usepackage{multirow}
\usepackage{float}
\usepackage{lineno}
\usepackage{xspace}
\usepackage[usenames,dvipsnames]{color}

\newcommand{\bef}{\begin{figure}}
\newcommand{\eef}{\end{figure}}
\newcommand{\bc}{\begin{center}}
\newcommand{\ec}{\end{center}}

\newcommand{\be}{\begin{equation}}
\newcommand{\ee}{\end{equation}}
\newcommand{\bea}{\begin{eqnarray}}
\newcommand{\eea}{\end{eqnarray}}

\begin{document}
\title{Role of Multi-Parton Interactions on $J/\psi$ production in $p+p$ collisions at LHC Energies}
\author{Dhananjaya Thakur}
\author{Sudipan De}
\author{Raghunath Sahoo}
\email{Raghunath.Sahoo@cern.ch}
\affiliation{Discipline of Physics, School of Basic Sciences, Indian Institute of Technology Indore, Simrol, Indore 453552, INDIA}

\author{Soumya Dansana}
\affiliation{Department of Physical Sciences, Indian Institute of Science Education and Research, Kolkata-741246, India}
\begin{abstract}
The production mechanism of quarkonia states in hadronic collisions is still to be understood by the scientific community. In high-multiplicity $p+p$ collisions, Underlying Event (UE) observables are of major interest. The Multi-Parton Interactions (MPI) is a UE observable, where several interactions occur at the partonic level in a single $p+p$ event. This  leads to dependence of particle production on event multiplicity. If the MPI occurs in a harder scale, there will be a correlation between the yield of quarkonia and total charged particle multiplicity. The ALICE experiment at the Large Hadron Collider (LHC) in $p+p$ collisions at $\sqrt{s}$ = 7 and 13 TeV has observed an approximate linear increase of relative $J/\psi$ yield, ($\frac{dN_{J/\psi}/dy}{<dN_{J/\psi}/dy>}$) with relative charged particle multiplicity density, ($\frac{dN_{ch}/dy}{<dN_{ch}/dy>}$). In our present work we have performed a comprehensive study of the production of charmonia as a function of charged particle multiplicity in $p+p$ collisions at LHC energies using pQCD-inspired multiparton interaction model, PYTHIA8 tune 4C, with and without Color Reconnection (CR) scheme. A detail multiplicity and energy dependent study is performed to understand the effects of MPI on $J/\psi$ production. The ratio of $\psi(2S)$ to $J/\psi$ is also studied as a function of charged particle multiplicity at LHC energies.   
\end{abstract}
\date{\today}
\maketitle

\section{Introduction}
\label{intro}
Understanding the production mechanisms of quarkonia states in proton-proton collisions is of great challenge for existing theoretical models. Various theoretical models such as the Color Singlet, non-relativistic QCD approach (NRQCD) and the Color Evaporation Model try to explain the heavy resonance states produced in hard processes~\cite{Brambilla,Lansberg}. The recent theoretical works~\cite{Butenschoen1,Butenschoen2,YMa} are dedicated in understanding the production cross section and polarization of $J/\psi$ by taking the inputs from recent LHC measurements~\cite{Aamodt1,Aaij,Aad,Khachatryan}. 
In high energy $p+p$ collisions the total event multiplicity can have a substantial contribution from Multi-Parton Interactions (MPI)~\cite{Sjöstrand, Bartalini}, which is an Underlying Event observable. The sum of all the processes that build up the final hadronic state in a collision is referred as the Underlying Event (UE). The Underlying Event includes fragmentation of beam remnant, multi-partonic interactions, and initial and final state radiation (ISR/FSR) associated with each interaction. In MPI, several interactions at the partonic level occur in a single $p+p$ collision. This leads to a strong dependence of particle production on total event multiplicity. MPI are commonly used to describe the soft underlying events such as the production of light quark and gluons. But it is observed that it can also contribute on the hard and semi-hard scale such as the production of particles containing heavy quarks like $J/\psi$, open heavy flavor etc. This contribution becomes more and more prominent with increasing energy~\cite{Drescher}. An early study of NA27 experiment reported that the charged particle multiplicity distributions are affected by the underlying events with open charm production~\cite{NA27}. This indicates a correlation between the yield of quarkonia and the total charged particle multiplicity. According to Ref.~\cite{Frankfurt}, the probability of MPI  increases towards smaller impact parameters. So the multiplicity dependence study in $p+p$ collisions is very interesting to understand the MPI effects on quarkonia production.   

Recently, ALICE experiment has observed an approximately linear increase of relative $J/\psi$ yields $\frac{dN_{J/\psi}/dy}{<N_{J/\psi}/dy>}$ as a function of relative charged particle multiplicity density $\frac{dN_{ch}/dy}{<dN_{ch}/dy>}$~\cite{Abelev:2012rz}. The QCD inspired model PYTHIA6 could not describe the results and shows an exactly opposite behaviour. An updated version of PYTHIA has been proposed, known as PYTHIA8, where Multiple-Parton Interaction (MPI) plays an important role in production of heavy quarks like charm and beauty. PYTHIA8 describes the increasing trend of open charm and non-prompt $J/\psi$ production as a function of charged particle multiplicity measured by ALICE  experiment at $\sqrt{s}$ = 7 and 13 TeV~\cite{Adam:2015ota, ALICE13TeV}. The Color Reconnection (CR) is an important mechanism to describe the interactions that can occur between colored fields during the hadronization process. CR is expected to occur at a significant rate at LHC energies because of high number of colored partons from both MPI and parton showers. CR is the most important ingredient to the UE contributions and hence plays an important role in the interplay between soft and hard processes. As there is no first principle calculation, the only way is to study it's effect via realistic models. It is found that PYTHIA8 describes the $<p_{T}>$ as a function of charged particle multiplicity by including the MPI with-CR \cite{Kar:2017shp} as well as explains the flow like pattern in $pp$ collisions~\cite{Ortiz:2013yxa}. In this work, we have studied the effect of MPI with and without-CR scheme on $J/\psi$ production as a function of charged particle multiplicity using 4C tuned~\cite{Corke:2010yf} PYTHIA8 event generator at different LHC energies. In the present work, for the first time we have studied the energy dependence of the event activity of $J/\psi$, which can provide more insight into the processes in an event in hadronic collisions.   

The paper is organised as follows. Event generation and Analysis methodology is described in Section~\ref{eventgen}. The results are discussed in Section~\ref{result}. The Section III is divided into three subsections. \hyperref[muldep]Multiplicity dependence of $J/\psi$ production is discussed in subsection A. \hyperref[energydep]In subsection B, we show the energy dependence behaviour and in  \hyperref[ratio]subsection C, we study the ratio between higher resonance state of $J/\psi$ ($\psi(2S)$) and $J/\psi$ as function of charged particle multiplicity. Finally in Section~\ref{sum}, we summarize our work. 
   

\section{Event generation and Analysis methodology}
\label{eventgen}
PYTHIA8 is the advanced version of PYTHIA6 coded in C++. One of the major improvements in PYTHIA8 with respect to PYTHIA6 is the implementation of MPI scenario, where $c$ and $b$ quarks can be produced via $2\rightarrow2$ hard sub-processes. The heavy-flavor production in PYTHIA8 includes the following main processes: (i) initial $c$ or $b$ quarks originate via first hardest $2\rightarrow2$ partonic interactions (e.g: $q\bar{q}\rightarrow \bar{c}c$, $g\bar{g}\rightarrow \bar{c}c$) and also has the finite production probability from the subsequent hard processes in MPI via the same mechanism, (ii) heavy quarks from gluon splitting hard processes (e.g: $g\rightarrow c\bar{c}$), and (iii) these gluons may originate from initial and final state radiation (ISR/FSR). Detailed explanation on PYTHIA8.2 physics processes can be found in Ref.~\cite{pythia8html}. 
 At the LHC energies, it is possible to have multiple parton-parton interaction (MPI) in $p+p$ collisions. Many of the models have included it in different ways to explain the LHC $p+p$ data~\cite{Srivastava:2017bcm,Drescher:2000ha,Kopeliovich:2013yfa}. In our present study we have used 4C tuned PYTHIA8.2. We have included varying impact parameter (MultipartonInteractions:bProfile=3) to allow all incoming partons to undergo hard and semi-hard interactions as well. We have used the MPI-based scheme of Colour Reconnection (ColourReconnection:mode(0)) of PYTHIA8.2. In this scheme, the produced partons are classified by the MPI system that undergoes a reconnection where partons from lower $p_{T}$ MPI systems are added to the dipoles defined by the higher $p_{T}$ MPI system such a way that minimizes total string length. This model of CR describes the underlying event observables very well compared to the other models as is reported by the ATLAS Collaboration in Ref.~\cite{atlaspubnote}. More details on various models of CR included in PYTHIA8.2 and their performances with respect to experimental data can be found in Refs.~\cite{Bierlich:2015rha, Abelev:2013bla}. After colour reconnection, all the produced partons, connected with strings, fragment into hadrons via the Lund String Model~\cite{Sjostrand:2006za, Andersson:2002ap}.
 
In this study, we have simulated inelastic, non-diffractive component of the total cross section for all hard QCD processes (HardQCD:all=on), which includes the production of heavy quarks. A $p_{T}$ cut of 0.5 GeV/c (using PhaseSpace:pTHatMinDiverge) is used to avoid the divergences of QCD processes in the limit $p_{T}\rightarrow0$. $J/\psi$ production is measured in the dimuon channel. We have specifically decayed $J/\psi$ to the dimuon channel and measured the yield of the reconstructed particles by defining an external decay mode. 

\bef[ht]
 \bc
\includegraphics[scale=0.38]{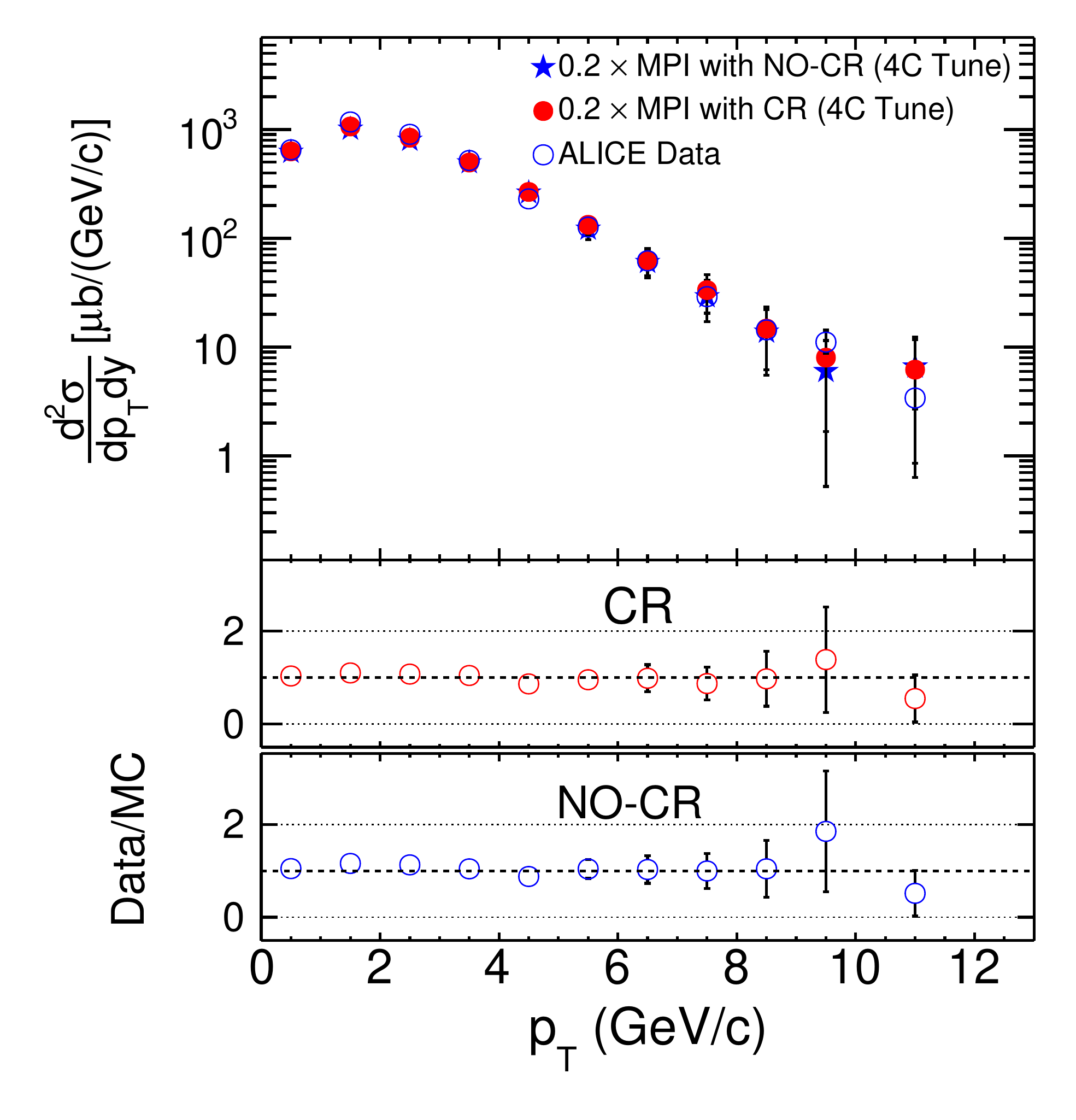}
\caption{ (Color online) Upper panel shows the comparison of ALICE data~\cite{Abelev:2012rz} and PYTHIA8 of $J/\psi$ production cross-section at $\sqrt{s}$ = 5.02 TeV. The open circles are ALICE data and solid circles and stars represent PYTHIA8 results with-CR and without-CR, respectively.  Lower panels show the ratio between data and PYTHIA8 for both CR and NO-CR cases.}
 \label{fig1}  
 \ec
 \eef
 
 \bef[ht]
 \bc
\includegraphics[scale=0.38]{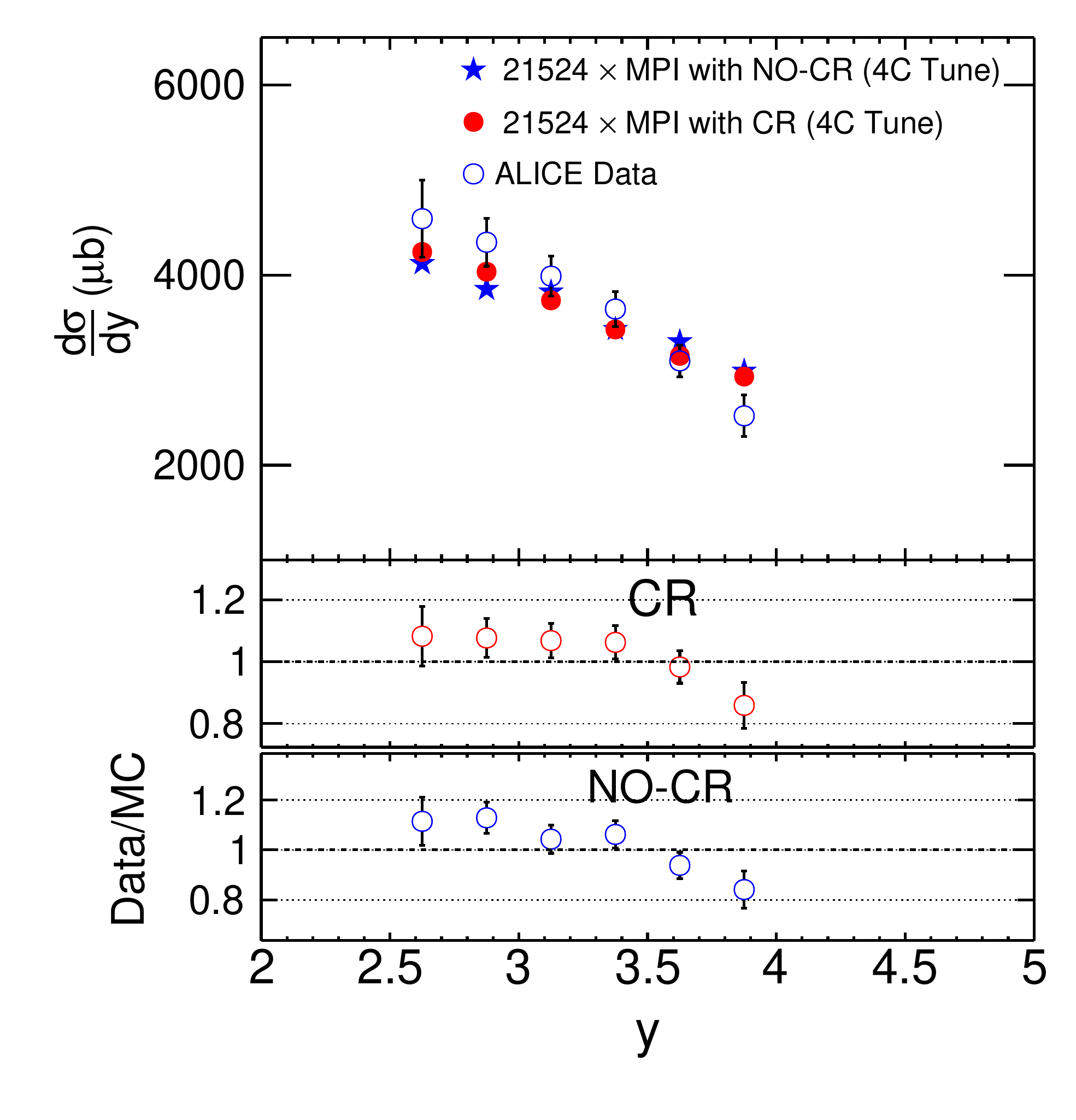}
\caption{ (Color online) Upper panel shows the comparison of ALICE data~\cite{Abelev:2012rz} and PYTHIA8 for d$\sigma$/dy of $J/\psi$ at $\sqrt{s}$ = 5.02 TeV. The open circles are ALICE data and solid circles and stars represent the PYTHIA8 results with-CR and without-CR, respectively.  Lower panels show the ratio between data and PYTHIA8 for both CR and NO-CR cases.}
 \label{fig2}  
 \ec
 \eef

\bef[ht]
 \bc
 \includegraphics[scale=0.38]{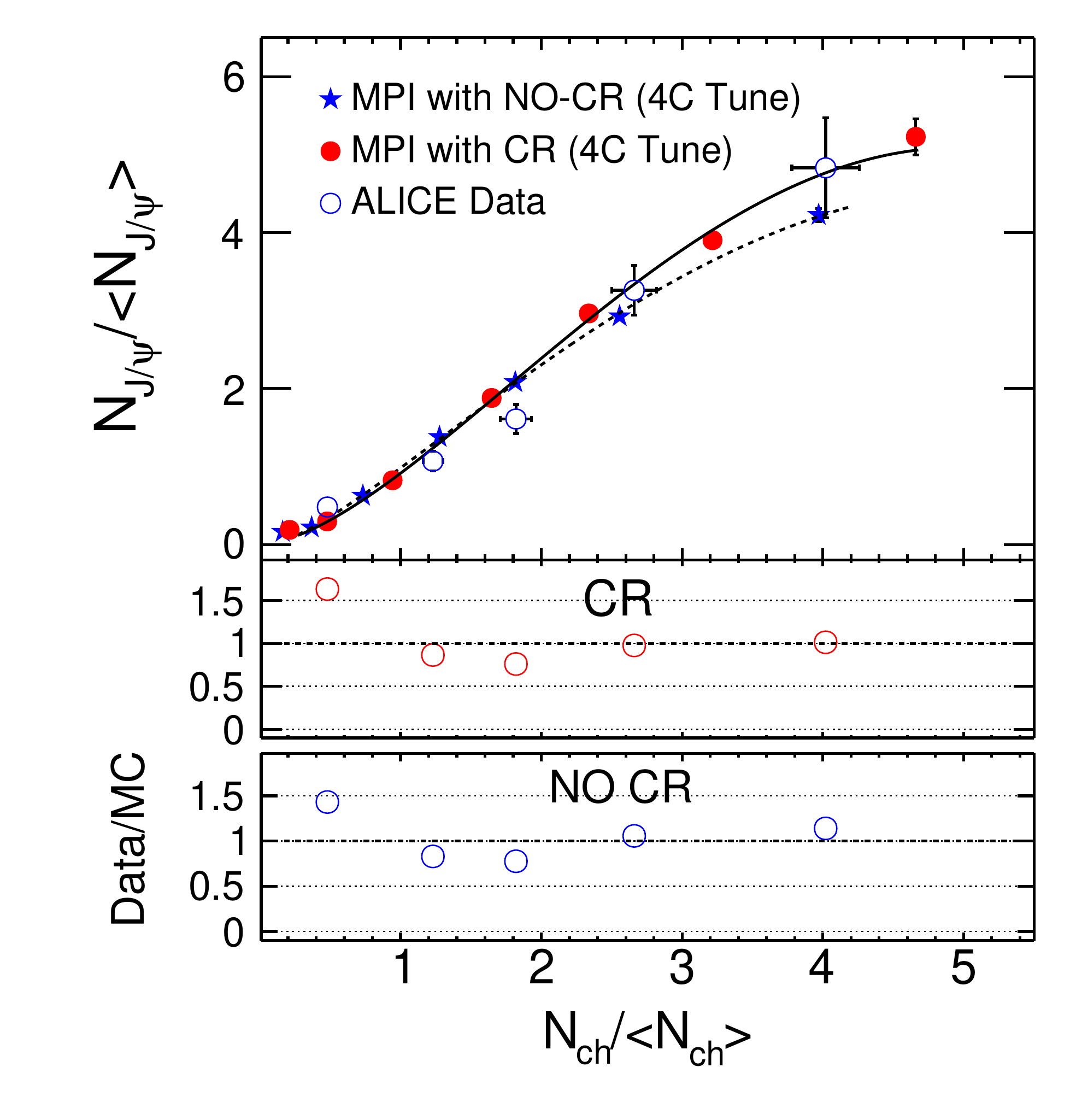}
\caption{ (Color online) Upper panel shows the comparison of ALICE data~\cite{Abelev:2012rz} and PYTHIA8 of relative $J/\psi$ yield ($N_{J/\psi}/<N_{J/\psi}>$) as a function of the relative charged particle multiplicity ($N_{ch}/<N_{ch}>$) for $\sqrt{s}$ = 7 TeV at forward rapidities. The open circles are ALICE data and solid circles and stars represent the PYTHIA8 results with-CR and without-CR, respectively. The lines are the fitted curves using the function as described in Eq.~\ref{eq4}. Lower panels show the ratio between data and PYTHIA for both CR and NO-CR cases.}
 \label{fig3}  
 \ec
 \eef
 
We have generated 100 million events for $p+p$ collisions at $\sqrt{s}$ = 0.9, 2.76, 5.02, 7 and 13 TeV using the option with-CR and without-CR scheme available in PYTHIA8.2. $J/\psi$ are reconstructed via dimuon channel ($J/\psi \rightarrow \mu^{+} + \mu^{-}$). The charged particle multiplicity yield, which is defined as $N_{ch} / < N_{ch} >$, is measured at midrapidity ($ |y| < 1.0$), where $N_{ch}$ is the mean of the charged particle multiplicity in a particular bin and $<N_{ch}>$ is the mean of the charged particle multiplicity in minimum-bias events. The relative $J/\psi$ yield is measured in forward rapidity ($2.5 <  y < 4.0$) using the following relation:
\bea
\frac{Y_{J/\psi}}{<Y_{J/\psi}>}= \frac{N_{J/\psi}^{i}}{N_{J/\psi}^{total}}\frac{N_{evt}}{N_{evt}^{i}},
\label{eq1}
\eea
where, $N_{J/\psi}^{i}$ and $N_{evt}^{i}$ are the number of $J/\psi$ and number of events in $i^{th}$ multiplicity bin, respectively. $N_{J/\psi}^{total}$ and $N_{evt}$ are the total number of $J/\psi$ produced and total number of minimum-bias events, respectively. As the frequency of lower multiplicity events is higher, the bin width is taken smaller at lower multiplicity and then subsequently higher to maximize the statistics at high multiplicity bins. To have a direct comparison of the obtained results with experimental data, charged particle multiplicity bins are chosen according to the available experimental measurements~\cite{Abelev:2012rz}.

%
%
%
%
%

 The statistical uncertainties are calculated in each multiplicity bin for both relative charged particle multiplicity ($N_{ch}/<N_{ch}>$) and relative 
$J/\psi$ yield ($N_{J/\psi}/<N_{J/\psi}>$). Uncertainty in $N_{ch}$ measurement is given by the ratio of RMS value of the charged particle multiplicity and square root of the number of charged particles in that bin ($N_{RMS}/\sqrt{N_{bin}}$). The ratio between RMS value of the minimum-bias (MB) charged particle multiplicity and square root of the number of minimum-bias charged particles ($N_{RMS}^{MB}/\sqrt{N_{MB}}$) gives the uncertainty in $<N_{ch}>$. The uncertainty to measure the number of $J/\psi$ particles is simply $\sqrt{N_{J/\psi}}$. These uncertainties are propagated using standard error propagation formula to estimate the uncertainties in relative charged particles multiplicity as well as in relative $J/\psi$ yield.
 
 \section{Results and Discussion}
 \label{result}

 To check the compatibility of PYTHIA8 with the experimental data we have compared $J/\psi$ production between data and PYTHIA8 in same kinematic range. Fig.~\ref{fig1} and Fig.~\ref{fig2} show the comparison of $J/\psi$ production cross-section as a function of $p_{T}$ and rapidity (y), respectively between data and PYTHIA for minimum-bias events. The open symbols represent the data obtained from ALICE and the solid circles and stars show the results from PYTHIA8 model at $\sqrt{s}$ = 5.02 TeV~\cite{Acharya:2017hjh}. The lower panels of the figures show the ratio between data and model for both CR and No-CR cases. It is observed that the PYTHIA8 explains the data very well. Similar study has been performed in other available LHC energies. We found a maximum of 60\% deviation of MC $p_T$ spectra from data for 13 TeV in certain  $p_T$ bins except for two low-$p_T$ bins ($p_T \sim 1$ GeV/c), where deviation is larger.  For the rapidity spectra, the deviation of MC from experimental data is less than 1\%. Similarly, for 7 TeV, 5.02 TeV and 2.76 TeV, the maximum deviation of ( $p_T$, y)-spectra are (50\%, 1\%), (1\%, 1\%) and (50\%, 1\%), respectively for few $p_T$ and rapidity bins. In most of the $p_T$ bins, the deviation is around 10-20\%, whereas rapidity spectra are very well reproduced by MC. These measurements provide us the confidence to study the quarkonia production using PYTHA8 in $p+p$  collisions at LHC energies.

Figure~\ref{fig3} shows the relative $J/\psi$ yield as a function of charged particle multiplicity for $\sqrt{s}$ = 7 TeV. The open circles show the results from ALICE experiment~\cite{Abelev:2012rz}. The star markers and solid circles represent our measurement with PYTHIA8. It is observed that 4C tuned PYTHIA8 with-CR and without-CR qualitatively reproduce the ALICE results. In the experimental data there were statistical and systematic uncertainties. We have added both the errors in quadrature and put single error bar in each data point. The PYTHIA8 data points have statistical uncertainties which are obtained using the method described in the previous section. Direct comparison of data to PYTHIA is not possible as the $N_{ch} / <N_{ch}>$ values are different for different cases. So, we have fitted the results obtained from PYTHIA8 model using a percolation inspired function~\cite{Ferreiro} as described in next section and interpolated the relative $J/\psi$ yield at the experimental $N_{ch} / <N_{ch}>$ values to make the ratio between data and PYTHIA. From the ratio it is clear that PYTHIA8 explains the data well for both CR and NO-CR cases.

The color reconnection is a final state effect, which takes place in $p+p$ collisions as implemented in PYTHIA8~\cite{pythia8html}. In our current studies, we use the MPI-based color reconnection, and it is observed that the difference in number of $J/\psi$ in CR and NO-CR is less than one per event. However, in highest two multiplicity bins it exceeds one. A quantitative study is performed in the next section. Therefore, one can conclude that the final state effect (CR) has little contribution to $J/\psi$ production. Although both CR and NO-CR describe experimental data well within the uncertainties, it is found that towards the higher multiplicity bins, MPI with-CR seems to reproduce the data better than that of MPI with NO-CR case. 

To explore the effect of MPI on quarkonia production with collision energies, we extend the analysis using PYTHIA8 to other available energies at the LHC.
 
 \subsection{Multiplicity dependence of $J/\psi$ production}
  \label{muldep}
 Figure~\ref{fig4} shows the relative $J/\psi$ yield as a function of relative charged particle multiplicity at forward rapidity ($2.5 <  y < 4.0$) for 4C tuned with and without-CR at $\sqrt{s}$ = 0.9, 2.76, 5.02, 7 and 13 TeV using the PYTHIA8, 4C tuned simulated data. The $J/\psi$ are reconstructed via dimuon channel. However, The ALICE results~\cite{Abelev:2012rz, ALICE13TeV} reveal that relative $J/\psi$ yield as a function of charged particle multiplicity at $\sqrt{s}$ = 7 TeV  gives very similar results for both dimuon and dielectron channels except the highest multiplicity bin. 
\bef[ht]
\bc
 \includegraphics[scale=0.42]{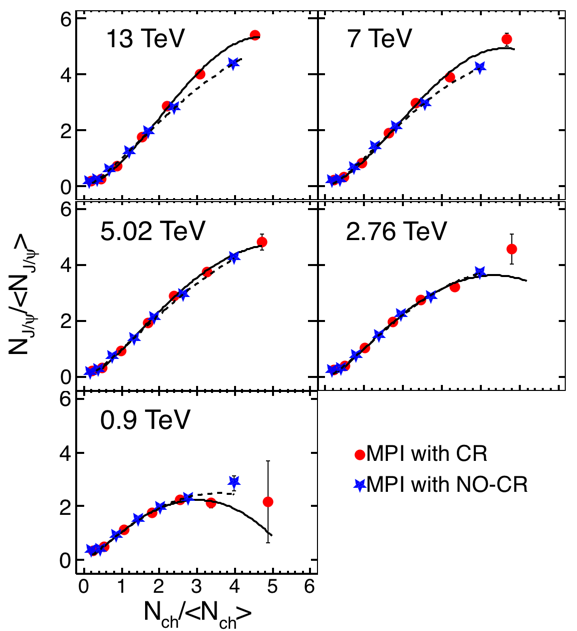}
 \caption  {(Color online) Relative $J/\psi$ yield as a function of relative charged particle multiplicity at forward rapidities at $\sqrt{s}$ = 0.9, 2.76, 5.02, 7 and 13 TeV using PYTHIA8. The circles and stars show the results with-CR and without-CR, respectively. The dotted line is fitted to MPI with NO-CR using Eq.~\ref{eq4}, where as the solid line represents the fitting to MPI with-CR using Eq.~\ref{eq4}. }
 \label{fig4}  
 \ec
 \eef
 
 \bef[ht]
 \bc
 \includegraphics[scale=0.28]{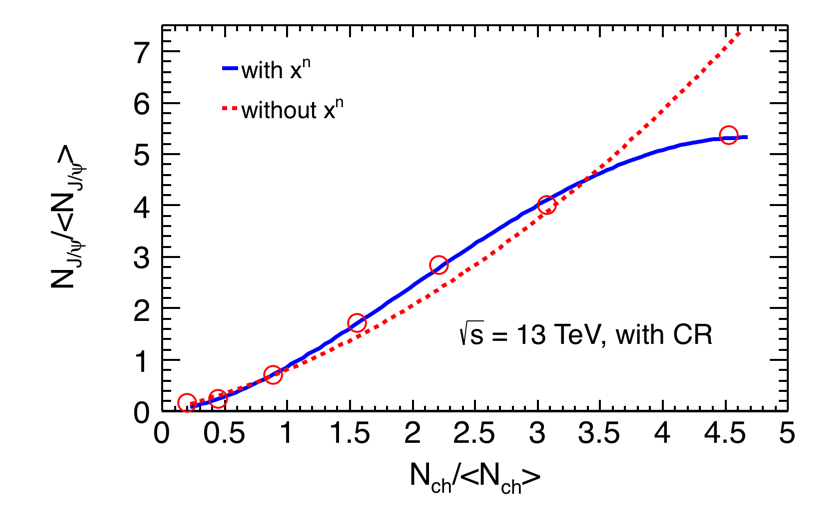}
 \caption  {(Color online) Relative $J/\psi$ yield as a function of the relative charged particle multiplicity for $p+p$ collisions at $\sqrt{s}$ = 13  using PYTHIA8 with-CR. The phenomenological function given by Eqn.~\ref{eq4} without the term $x^{n}$ is fitted to understand the validity of string percolation. This is shown by the dotted line. The solid line shows the fitting using  Eq.~\ref{eq4}, which has an additional term, $x^{n}$ over the string percolation function. This explains a possible saturation effect towards higher multiplicities.} 
\label{fig5} 
 \ec
 \eef
 \bef[ht]
 \bc
 \includegraphics[scale=0.36]{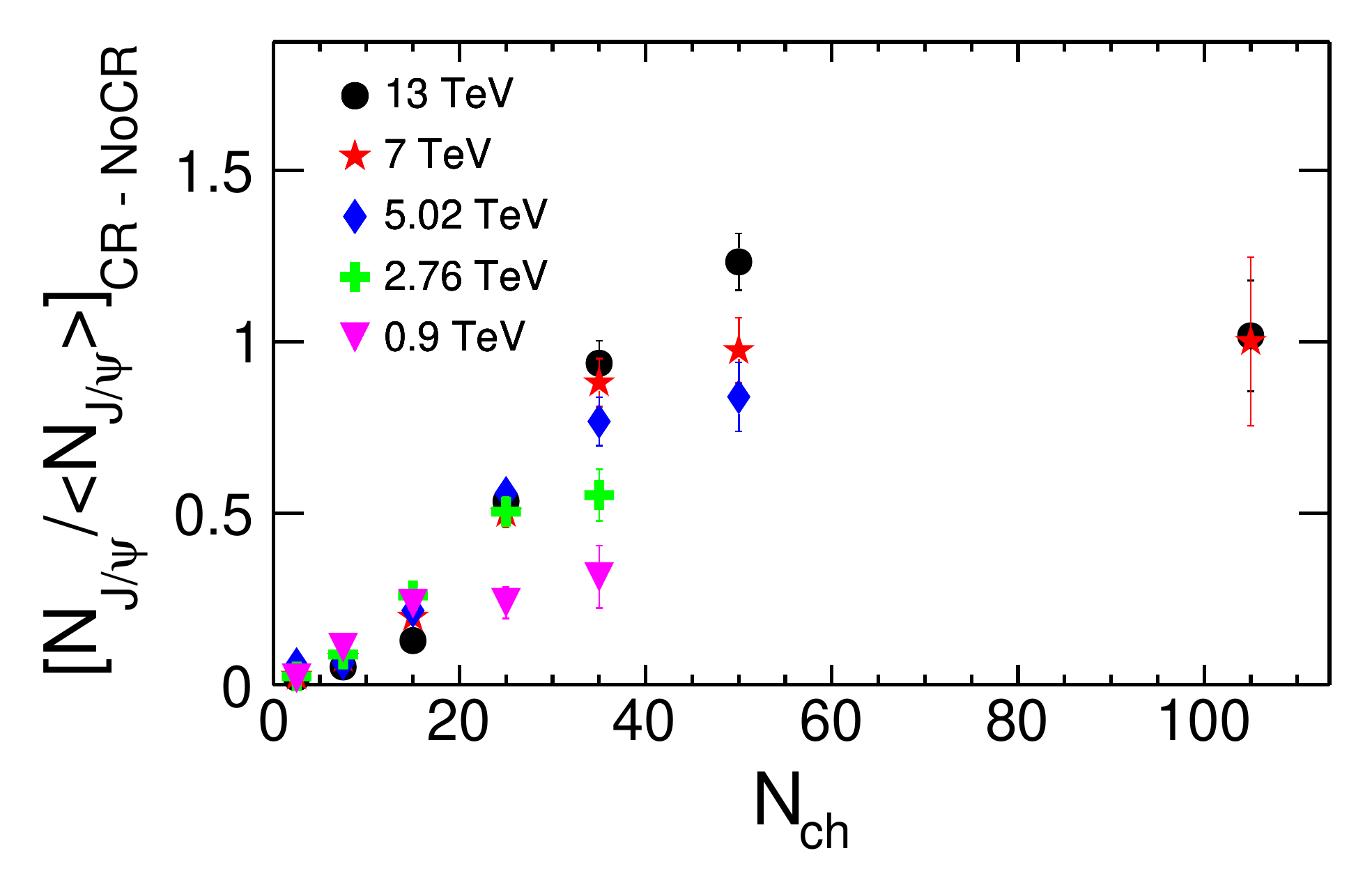}
 \caption {(Color online) Difference of relative $J/\psi$ yield from MPI with-CR to MPI without-CR as a function of the charged particle multiplicity at $\sqrt{s}$ = 0.9, 2.76, 5.02, 7 and 13 TeV using PYTHIA8.} 
\label{fig6} 
 \ec
 \eef

\begin{table*}[htb]
\centering
 \caption{(Color online) The difference values of relative $J/\psi$ with-CR and without-CR, ($N_{J/\psi}/<N_{J/\psi}>)_{\rm CR - No CR}$ for $\sqrt{s}$ = 0.9, 2.76, 5.02, 7 and 13 TeV.}
\small
\begin{center}
\begin{tabular}{|c|c|c|c|c|c|c|}
\hline
 $N_{ch}$ bin&   $\sqrt{s}$ =  0.9 TeV    &  $\sqrt{s}$ =  2.76 TeV   & $\sqrt{s}$ =  5.02 TeV  & $\sqrt{s}$ =  7 TeV & $\sqrt{s}$ =  13 TeV      \\ \hline
  0-5      &0.019 $\pm$ 0.026   & 0.024 $\pm$ 0.016    & 0.059 $\pm$0.013  &0.025 $\pm$ 0.013 &0.020 $\pm$ 0.011    \\ \hline
  5-10    &0.108 $\pm$ 0.018   & 0.088 $\pm$0.012    &0.063 $\pm$ 0.011   &0.077 $\pm$ 0.010 &0.052 $\pm$ 0.009        \\ \hline
  10-20   &0.236 $\pm$ 0.025  &0.260 $\pm$0.020     &0.217 $\pm$ 0.019  &0.197 $\pm$ 0.018 &0.129 $\pm$ 0.017       \\ \hline
  20-30   &0.239 $\pm$ 0.046  &0.505 $\pm$ 0.042     &0.558 $\pm$ 0.041  &0.500 $\pm$ 0.041  &0.535 $\pm$ 0.038         \\ \hline
  30-50   &0.314 $\pm$ 0.091  &0.553 $\pm$ 0.075     &0.768$\pm$ 0.070    &0.881 $\pm$ 0.070  &0.938 $\pm$ 0.066          \\ \hline
  50-100   &         -               &            -                &0.840 $\pm$ 0.101  &0.975 $\pm$ 0.095  &1.233 $\pm$ 0.083         \\ \hline
  100-150  &        -               &           -                 &          -              &1.001 $\pm$ 0.246 &1.017 $\pm$ 0.161         \\ \hline
\end{tabular}
\end{center}
\label{tablefin}
\label{table1}
\end{table*}

The circles show the results with-CR scheme and the stars show the results without-CR. It is observed that the relative $J/\psi$ yield is linearly increasing with charged particle multiplicity for all the centre-of-mass energies for both with and without-CR cases. We have fitted the results with a percolation inspired function~\cite{Ferreiro}. It is found that the function itself can not describe the results well unless we add an extra term $x^{n}$ in the function:  
\bea
\frac{Y_{J/\psi}}{<Y_{J/\psi}>}= A[Bx + Cx^{2} + Dx^{n}],
\label{eq4}
\eea
where A, B, C, D and n are the parameters and $x = N_{ch}/<N_{ch}>$. The solid lines show the fittings with-CR and the the dashed lines show the fittings without-CR. The difference between solid line and dashed line increases with higher multiplicity bins as well as with higher centre-of-mass energies, in particular, at $\sqrt{s}$ = 7 and 13 TeV. This indicates that the CR effects on $J/\psi$ production is more prominent at high multiplicity region of higher centre-of-mass energies.

Till now, the percolation theory has been successful in describing the relative $J/\psi$ yield as a function of relative charged particle multiplicity~\cite{Ferreiro}. 
To see if a string percolation scenario of particle production in $p+p$ collisions is valid at the highest available energy, {\it i.e.} at $\sqrt{s}$ = 13 TeV, we take
PYTHIA8 simulated data in our current study.  We compare string percolation model expectations using the function ($f(x) = Ax + Bx^{2}$) and found a deviation at
large multiplicity bins, where a kind of saturation behaviour is seen in PHYTHIA8. An additional term in the above equation, $x^{n}$ seems to describe the behaviour. 
This is shown in Fig.~\ref{fig5}. This study shows that the increase of $J/\psi$ yield as a function of charged particle multiplicity from PYTHIA simulated data doesn't
seem to follow exactly like a percolation type of behaviour rather a tendency to saturate at high multiplicity domain. The $p+p$ experimental data at other LHC energies
 would be very useful in understanding the particle production at higher multiplicities, helping to validate the percolation theory and/or tuning
the PYTHIA8 event generator with new physics inputs.

To understand the CR effects on $J/\psi$ production quantitatively, we have subtracted the yield of relative $J/\psi$ production with no CR from the yield with-CR and plotted as a function of charged particle multiplicity for different centre-of-mass energies, which is shown in Fig.~\ref{fig6}. It is found that the difference of relative $J/\psi$ yield between with-CR and without-CR increases as a function of charged particle multiplicity as well as with increase of centre-of-mass energy. The excess values of relative $J/\psi$ are shown in the Table~\ref{table1} for different multiplicity bins at all the discussed energies. We could not perform our studies at lower centre-of-mass energies for higher multiplicity bins due to higher statistical fluctuations. Although the numbers are very small, they are very significant as the $J/\psi$ yield is measured in event-by-event basis and normalised with total MB events using Eq. 1. It should be noted that the relative $J/\psi$ yield is the excess of $J/\psi$ production per event in a given multiplicity bin with respect to $J/\psi$ per event in MB.    
\bef[ht]
\bc
\includegraphics[scale=0.28]{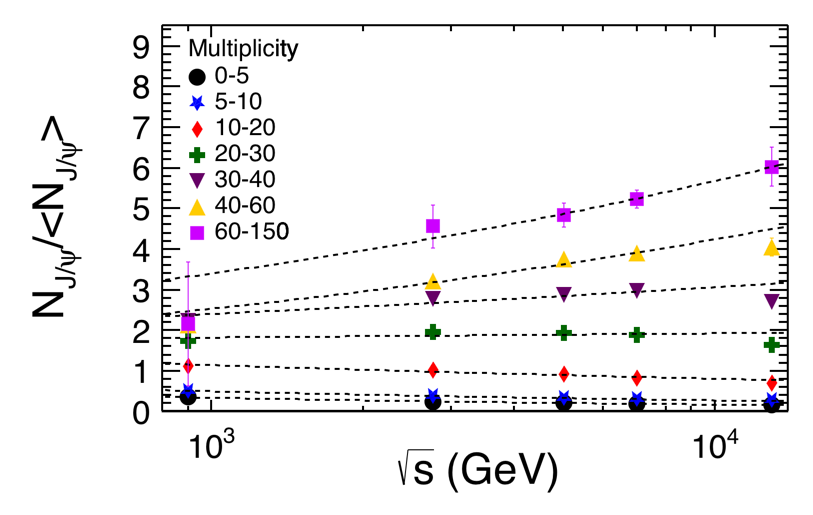}
\caption {(Color online) Relative $J/\psi$ yield  as a function of centre-of-mass energy, ($\sqrt{s}$) using PYTHIA8 with-CR. Different symbols are for different charged particle multiplicity bins. The dashed lines are the phenomenological fitting using the function, $y=Ax^n$.}
 \label{fig7}  
 \ec
 \eef
 
It is expected that CR occurs in a significant rate at the LHC energies due to higher number of colored partons from MPI. In our current study, we have used the default MPI based CR. Here, partons from lower $p_{T}$ MPI are merged with the ones at higher $p_{T}$ MPI~\cite{Cuautle:2016ukm}. At LHC energies due to high density of colored partons, there is a substantial degree of overlap of many colored strings in the position and momentum phase space. Therefore, there is higher probability of color reconnection. The partons from two different MPIs can reconnect via color strings with the minimization of the string length. This study reveals that with increase of MPI, the probability of color reconnection increases and hence the probability to combine charm and anti-charm quark becomes higher thereby producing  higher number of $J/\psi$ particles.  
 
  \bef[ht]
 \bc
 \includegraphics[scale=0.28]{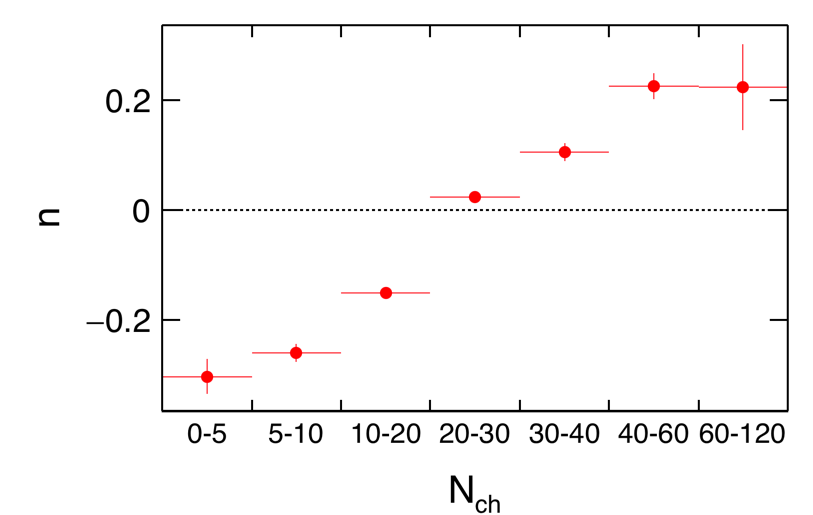}
 \caption{(Color online) The fitting parameter, $n$ for different multiplicity bins.}
 \label{fig8}  
 \ec
 \eef
 \subsection{Energy dependence of $J/\psi$ production}
 \label{energydep}
 As experimental data at different energies are not yet available, it is worth studying the effect of MPI and CR on charmonia production at various collision energies
 using PYTHIA8. Figure~\ref{fig7} shows the relative $J/\psi$ yield as a function of centre-of-mass energy for different charged particles multiplicity bins using PYTHIA8 with-CR. It is 
 observed that for higher multiplicity bins the relative $J/\psi$ production increases with energy and the trend changes as we go to lower multiplicity bins. To get a quantitative idea, we have fitted the results with a phenomenological function, $y=Ax^{n}$, where $A$ and $n$ are the parameters. The parameter, $n$ indicates the rate of increase of relative $J/\psi$ as a function of centre-of-mass energy for a particular multiplicity bin. Figure~\ref{fig8} shows $n$ as a function of different multiplicity bins. It is found that $n$ increases with the increase in charged particle multiplicity ($N_{ch}$). The values of $n$ are negative up to 10-20 multiplicity bins and becomes positive towards higher multiplicity bins. This indicates that MPI effects dominate for $J/\psi$ production for $N_{ch} >$ 20. This brings a 
threshold number for charged particle multiplicity in the final state for $p+p$ collisions in order to observe substantial MPI effects on charmonia production.  

\subsection{Ratio of $\psi(2S)$ and  $J/\psi$}
\label{ratio}
As $\psi(2S)$ is the first excited state of $J/\psi$, we have made an attempt to study the ratio of both as a function of charged particle multiplicity for different
collision energies. Same procedure for the reconstruction of $\psi(2S)$ is followed as is done for $J/\psi$, which is discussed in previous sections.
Figure~\ref{fig9} shows the multiplicity dependence of the ratio of relative yield of $\psi(2S)$ to $J/\psi$ using PYTHIA8 MPI with-CR. It is interesting to see that 
the ratio is independent of collision energy as well as multiplicity within statistical uncertainties.  The energy independent behaviour of the ratio of relative yields 
of $\psi(2S)$ to $J/\psi$ is also supported by color evaporation models~\cite{Bhaduri:2008zs}. At LHC energies there are sufficient energies to produce $c\bar{c}$  and as the mass difference between 
the two charmonia states is smaller compared to the collision energy, the production probability would be similar for both the cases in each multiplicity bin. This could be the reason of energy and multiplicity independent behaviour of the ratio of relative yields of $\psi(2S)$ to $J/\psi$. The CMS experiment preliminary 

\bef[ht]
\bc
\includegraphics[scale=0.27]{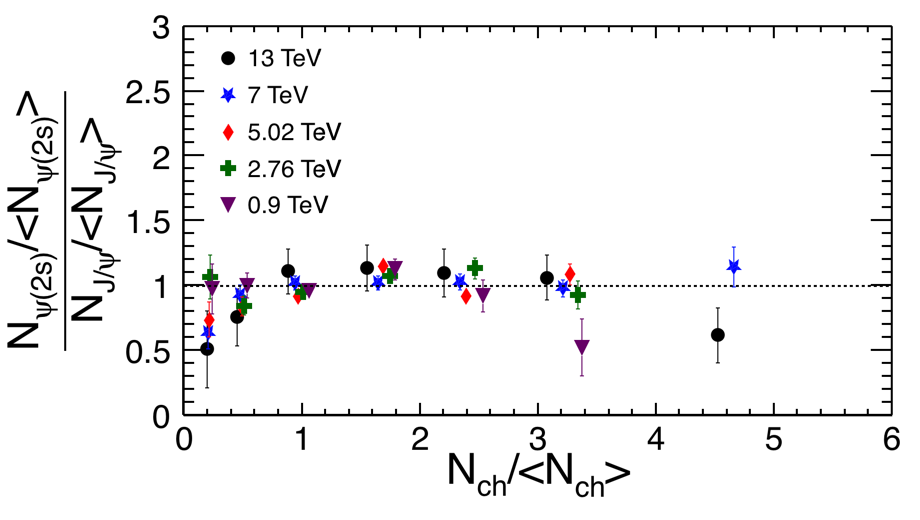}
\caption {(Color online) Ratio of the relative yields of $\psi(2S)$ and $J/\psi$ as a function of relative multiplicity for different centre-of-mass energies ($\sqrt{s}$)
using PYTHIA8 MPI with-CR.}
 \label{fig9}  
 \ec
 \eef         
 
results also show a similar trend at $\sqrt{s}$ = 7 TeV~\cite{Giovanni}. From Eq.~\ref{eq1} we can write the relative ratio of $\psi(2S)$ to $J/\psi$ as: 

\bea
\frac{Y_{\psi(2S)}/<Y_{\psi(2S)}>}{Y_{J/\psi}/<Y_{J/\psi}>}= \frac{N_{\psi}^{total}}{N_{J/\psi}^{total}}\frac{N_{J/\psi}^{i}}{N_{\psi}^{i}}
\label{eq5}
\eea

Here, the ratio $N_{\psi}^{total}/N_{J/\psi}^{total}$ is fixed for a particular collision energy. 
This implies that the $\psi(2S)$ and $J/\psi$ in each multiplicity bin are almost produced in same proportion within statistical uncertainties. This requires further investigations for a clear understanding of the underlying physics mechanism(s). The availability of experimental data in the near future will make the scenario more clear.

 \section{Summary}
  \label{sum}
 In summary, energy dependent study of $J/\psi$ production as a function of charged particle multiplicity has been performed using 4C tuned PYTHIA8 MC event generator. The relative $J/\psi$ are measured as a function of charged particle multiplicity in $p+p$ collisions at $\sqrt{s}$ = 0.9, 2.76, 5.02, 7 and 13 TeV. The $J/\psi$ are reconstructed via dimuon channel at forward rapidity ($2.5 <  y < 4.0$), whereas the charged particles are measured at mid-rapidity region ($ |y| < 1$). PYTHIA8 simulated data are generated using with-CR and without-CR. It is found that both  explain the ALICE data qualitatively for $p+p$ collisions at $\sqrt{s}$ = 7 TeV. A detailed quantitive study is performed to understand the CR effects on MPI for $J/\psi$ production. We observed that the difference between CR and No-CR increases with charged particle multiplicity as well as with centre-of-mass energy. The CR effects mostly dominate at high multiplicity bins for $\sqrt{s}$ = 7 and 13 TeV. The relative $J/\psi$ yield as a function of $\sqrt{s}$ shows a monotonic increase for higher multiplicity bins, whereas for bins with $N_{ch} \leq$ 20, it shows an
 opposite behaviour. This brings a threshold number for charged particle multiplicity in the final state for $p+p$ collisions in order to observe substantial MPI effects on charmonia production. This is a very important observation in view of the interesting properties shown by high multiplicity events in $p+p$ collisions at the LHC energies. The ratio of relative yield between $\psi(2S)$ and $J/\psi$ is measured at all centre-of-mass energies are presented as a function of relative charged particle multiplicity. $\psi(2S)$ to $J/\psi$ ratio shows both energy and multiplicity independent behaviour. It will be very interesting to get the experimental measurements, which can help to understand the production mechanism of charmonia states in $p+p$ collisions.   
 
\section{Acknowledgement} 
DT acknowledges UGC, New Delhi, Government of India for financial supports.  SD and RNS acknowledge the financial supports  from  ALICE  Project  No. SR/MF/PS-01/2014-IITI(G) of  Department  of  Science $\&$ Technology,  Government of India. Soumya Dansana would like to thank DST-INSPIRE program of Government of India for financial supports. This  research used  resources  of  the  LHC  grid  computing  centers  at
Variable Energy Cyclotron Center, Kolkata.

 
 
 \end{document}